\documentclass[sigconf]{acmart}

\usepackage{libertine}
\usepackage{color}
\usepackage{tabularx}
\usepackage{subfigure}
\usepackage{svg}
\usepackage{bm}
\usepackage{ulem}
\usepackage{algpseudocode}
\usepackage{algorithm,algorithmicx}
\usepackage{amsmath}
\usepackage{mathrsfs}
\usepackage{amsfonts}
\usepackage{hyperref}
\usepackage{enumitem}
\AtBeginDocument{%
  }

\setcopyright{acmcopyright}
\copyrightyear{2023}
\acmYear{2023}
\acmDOI{XXXXXXX.XXXXXXX}
\acmConference[NOSSDAV '23]{the 33rd edition of the Workshop on Network and Operating System Support for Digital Audio and Video}{June 7-10,
  2023}{Vancouver, Canada}
\acmPrice{15.00}
\acmISBN{978-1-4503-XXXX-X/18/06}

\begin{document}
\begin{sloppypar}
\title{Improving ABR Performance for Short Video Streaming Using Multi-Agent Reinforcement Learning with Expert Guidance}

\author{Yueheng Li}
\authornote{Both authors contributed equally to this research.}
\affiliation{%
  \institution{Nanjing University}
   \country{China}
  }

\author{Qianyuan Zheng}
\authornotemark[1]
\affiliation{%
  \institution{Nanjing University}
   \country{China}
  }

\author{Zicheng Zhang}
\affiliation{%
  \institution{Nanjing University}
   \country{China}
  }

\author{Hao Chen}
\authornote{Hao Chen is the corresponding author, \textit{chenhao1210@nju.edu.cn}.}
\affiliation{%
  \institution{Nanjing University}
   \country{China}
  }

\author{Zhan Ma}
\affiliation{%
  \institution{Nanjing University}
   \country{China}
  }


\renewcommand{\shortauthors}{Y. Li et al.}
%

\begin{abstract}
In the realm of short video streaming, popular adaptive bitrate (ABR) algorithms developed for classical long video applications suffer from catastrophic failures because they are tuned to solely adapt bitrates. Instead, short video adaptive bitrate (SABR) algorithms have to properly determine {\it which video} at {\it which bitrate level} together for content prefetching, without sacrificing the users' quality of experience (QoE) and yielding noticeable bandwidth wastage jointly. Unfortunately, existing SABR methods are inevitably entangled with slow convergence and poor generalization. Thus, in this paper, we propose Incendio, a novel SABR framework that applies Multi-Agent Reinforcement Learning (MARL) with Expert Guidance to separate the decision of video ID and video bitrate in respective buffer management and bitrate adaptation agents to maximize the system-level utilized score modeled as a compound function of QoE and bandwidth wastage metrics. To train Incendio, it is first initialized by imitating the hand-crafted expert rules and then fine-tuned through the use of MARL. Results from extensive experiments indicate that Incendio outperforms the current state-of-the-art SABR algorithm with a 53.2\% improvement measured by the utility score while maintaining low training complexity and inference time.
\end{abstract}




\maketitle

\vspace{-7pt}
\section{Introduction}
\begin{figure*}[t]
    \setlength{\abovecaptionskip}{0.1cm}
    \centering
    \setlength{\belowcaptionskip}{-12pt}
    \includegraphics[width=\linewidth]{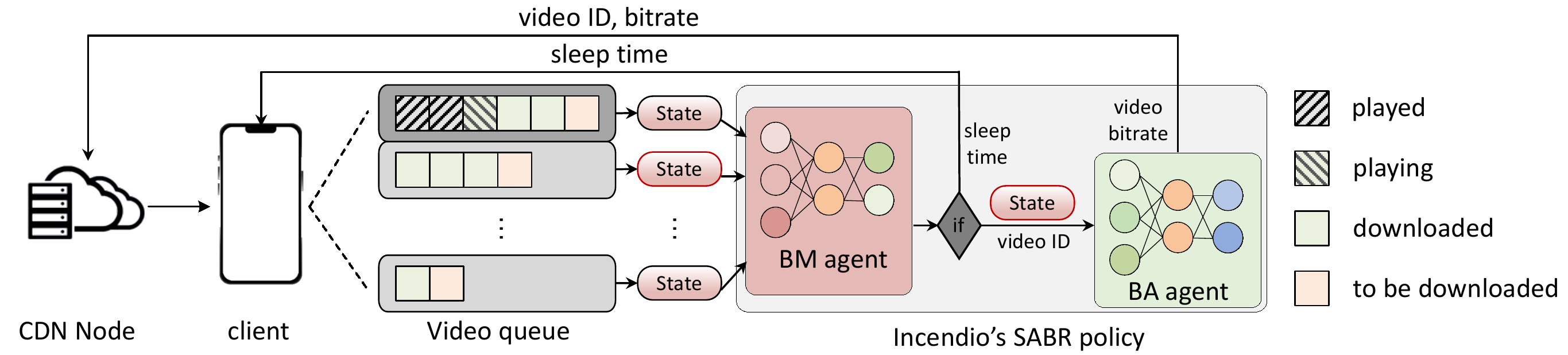}
    \caption{{\it Incendio} uses two hierarchical agents which are responsible for buffer management (BM-agent) and bitrate adaption (BA-agent) respectively. They make their decisions based on the observations including past throughput measurements, user retention rate, video chunk size, and buffer status at each decision iteration.}
    \label{fig:sys}
\end{figure*}

In recent years, there has been a significant surge in using short video streaming applications such as Kwai~\cite{kwai} and TikTok~\cite{tiktok} for entertainment, social connection, etc, resulting in exponential traffic growth. Such a short video service largely differs from traditional long video streaming scenarios like video-on-demand (VoD), in which it allows the user to promptly switch to his/her interested content by just scrolling the touch screen. To this end, we often need to prefetch personalized content into the local buffer properly.
Prefetching as many short videos as possible into the local buffer ensures the quality of experience (QoE) during consumption but often leads to significant bandwidth wastage. On the other hand,  inadequate buffering may cause noticeable start-up delays when scrolling to the next one that is not yet cached.
To tackle these challenges, content providers mainly resort to short video adaptive bitrate algorithms (SABR) for optimizing the user's QoE and reducing bandwidth wastage simultaneously, for which the SABR algorithm needs to determine which video to download or remain idle and then identify which bitrate of this specific video to preload. To this end, network conditions, client buffer status, chunk sizes, as well as the users' viewing preferences can be jointly leveraged to make a proper decision. A short video streaming grand challenge was  held in ACM Multimedia 2022 (MMGC2022~\cite{mmgc2022}) that attracted numerous competitive solutions. 

Both rules-based and reinforcement learning (RL) based SABR methods were developed. For example, PDAS~\cite{PDAS}, a typical rules-based approach, offers the leading performance in MMGC2022, in which it applies a probability-based reward function and a handcrafted buffer management model. However, rules-based approaches are often criticized for their poor generalization to different environments since fixed control rules could not thoroughly characterize system behaviors for all scenarios in practice~\cite{pensieve,Oboe}.  Additionally, as PDAS is a variant of model predictive control (MPC~\cite{MPC}) that uses a greedy heuristic search for decision-making, its decision inference time grows exponentially as the length of the optimization horizon increases.

Thus, RL-based approaches are introduced to overcome these issues through the use of neural networks to make a direct connection with environmental observation and action. 
For instance, DAM~\cite{DAM}, an RL-based SABR method~\cite{tang2020discretizing,fan2022soft}, makes decisions for buffer management and bitrate adaptation simultaneously based on the probability of every combination of the atomic actions.  During the training of DAM, it suffers from a slow convergence rate (and thus an extremely-long time duration) to the global optimality which is attributed to the search in large discrete action space  that is closely related to the number of videos in the queue and the total bitrate levels for each video.

This paper, therefore, proposes the Incendio, yet another novel SABR framework, to address the aforementioned issues in existing approaches for joint optimization of QoE and bandwidth efficiency.  We separate the decision of respective buffer management and bitrate adaption in a sequential manner, i.e., sub-task decomposition, to which the hierarchical multi-agent reinforcement learning (MARL) is devised to simultaneously train them to optimize a compound reward.  This greatly reduces the action space for optimality search,  accelerating neural network training with a much faster convergence rate. On the other hand, instead of executing the MARL from the scratch, we propose imitation learning to pre-train Incendio from a rudimentary state to an expert state by leveraging human experience, which further reduces the number of invalid trials in MARL and also mitigates the risk of sub-optimality.

We evaluate the performance of Incendio by comparing it against state-of-the-art algorithms including PDAS~\cite{PDAS}, MPC~\cite{MPC}, and DAM~\cite{DAM}, under various network and users' preference conditions (as detailed in §\ref{ssec:Methodology}). Our results indicate that Incendio consistently outperforms the existing algorithms across all scenarios. On average, Incendio exhibits a 53.2\% improvement to the award-winning PDAS under the measurement of overall utility score (as reported in §\ref{ssec:Results}), while maintaining exceptional training efficiency (as reported in §\ref{ssec:convergence}) and feasibility of deployment (as reported in §\ref{ssec:running_time}).

\vspace{-4pt}
\section{Backgrounds and Related Works}
\label{sec:backgrounds}
This section commences by first briefing the optimization objective function well-accepted in the context of the SABR problem. Subsequently, we review both rules-based and learning-based  SABR algorithms and discuss their limitations.

{\bf Optimization objective.} Unlike long video streaming applications (e.g., VoD) that mainly focus on enhancing the user's QoE, short video streaming has to consider QoE improvement and bandwidth efficiency (e.g., bandwidth wastage reduction) jointly.
Consequently, the optimization objective of SABR involves not only the QoE model~\cite{pensieve, MPC, Oboe} {but also a bandwidth usage penalty term, which is defined as the overall utility score~\cite{mmgc2022}}, i.e., 
\begin{equation}
\label{eq:score}
    \setlength{\belowdisplayskip}{-1pt}
    \begin{aligned}
        U_{i} & =Q o E_{i}-{Bandwidth}_{i} \\& =\sum_{m}\left(R_{m}-S_{m}\right)-\sum_{n} \mu \cdot T_{n}-\sum_{n} \nu \cdot bw_{n},
    \end{aligned}
\end{equation}
where ${m}$ and ${n}$ represent the index of played and downloaded chunks of video ${i}$. $R_{m}$ and $S_{m}$ respectively denote the quality (bitrate) and its fluctuation for each played chunk $m$. And $T_{n}$ and $bw_{n}$ respectively represent the rebuffering time and bandwidth usage caused by downloading chunk $n$. We set the coefficients $\mu=1.85, \nu=0.5$ as suggested in~\cite{mmgc2022} which are consistent with the other methods for a fair comparison. 

{\bf Rule-based SABR approaches.} 
APL~\cite{APL}  presented an adaptive preloading mechanism through the use of Lyapunov optimization to jointly maximize playback smoothness and minimize bandwidth waste. However, APL made a fixed bitrate assumption for short videos, which is  impractical for real-world applications. PDAS~\cite{PDAS} incorporates user retention rate\footnote{User retention rate indicates the percentage of the users that choose to continue the watching of current video by statistics, which can be provided by content providers at the granularity of chunk.} for more accurate QoE prediction and utilizes MPC rules to facilitate decision-making by comparing all possible combinations of future actions, which has attained state-of-the-art performance (ranked first in MMGC2022). Nevertheless, PDAS's hyperparameters are highly context-dependent, making the model hardly generalizable to various conditions with different user preferences and networks (refer to §\ref{ssec:Results} for further elaboration).

{\bf Learning-based approaches} have demonstrated their superiority in traditional ABR tasks~\cite{pensieve,MPC,comyco}. For SABR, LiveClip~\cite{liveclip} employs reinforcement learning to anticipate video switch events and dynamically modify preload orders, while overlooking the issues of bitrate adaptation and bandwidth conservation. DUASVS~\cite{DUASVS} utilizes integrated learning to develop a control policy for both decisions of prefetch threshold and video bitrate. DAM~\cite{DAM} achieves superior performance (ranked first among all learning-based techniques in MMGC2022) by incorporating the user retention rate into the reward function and minimizing training complexity through the utilization of action masks. However, the aforementioned learning-based approaches suffer from slow convergence in training, given a large discrete action space in SABR tasks which is derived by multidimensional decisions of whether to sleep or not, the video ID (to-be-prefetched), and bitrate level (refer to §\ref{ssec:convergence} for further elaboration).

\section{System Overview}
The system architecture of Incendio is illustrated in Figure~\ref{fig:sys}. Each short video is sliced into chunks with a length of 1s. Each chunk is encoded into several bitrate versions and stored in a content delivery network (CDN) node. The client downloads video chunks from the CDN node purposely and maintains a local buffer for each short video in the video queue, including the current playing video and several recommended videos. Different videos are marked with different IDs. Every time when the user scrolls the screen, the second video in the queue starts to play, and the downloaded but unplayed chunks for the previous video are cleared, resulting in a waste of bandwidth. In the meantime, a new video suggested by the video recommendation mechanism will be appended to the queue.

The RL agent of Incendio consists of two hierarchical agents responsible for buffer management (BM-agent) and bitrate adaption (BA-agent) respectively. For each decision iteration, the BM-agent chooses to sleep for a fixed duration or selects a video that needs the most buffering based on observations of past throughput measurements and the status for each video in the queue including user retention rate, remaining buffer size, average chunk size, rebuffering time, and bitrate as well as its fluctuation of last downloaded chunk. 
If BM-agent makes a sleep decision, the client's preloading process will be suspended for a predefined time duration. Otherwise, BM-agent decides which video to prefetch with a video ID. Subsequently, BA-agent determines the bitrate of the next chunk for this video to download based on video states and network conditions. Afterward, the client submits the request with a video ID and its bitrate to the CDN node and promotes a new round of interaction. 


The number of videos in the queue and the bitrate levels for each of them is determined by the underlying streaming platform. Here we adopt the same settings used in MMGC2022, which comprises five videos in the queue and three bitrate levels for each of them. Once the offline training is completed, the Incendio policy remains fixed for task inference. We detail the agent design and training method of Incendio in the following section.

\begin{figure}[t]
    \setlength{\abovecaptionskip}{0.1cm}
    \centering
    \setlength{\belowcaptionskip}{-17pt}
    \includegraphics[width=\linewidth]{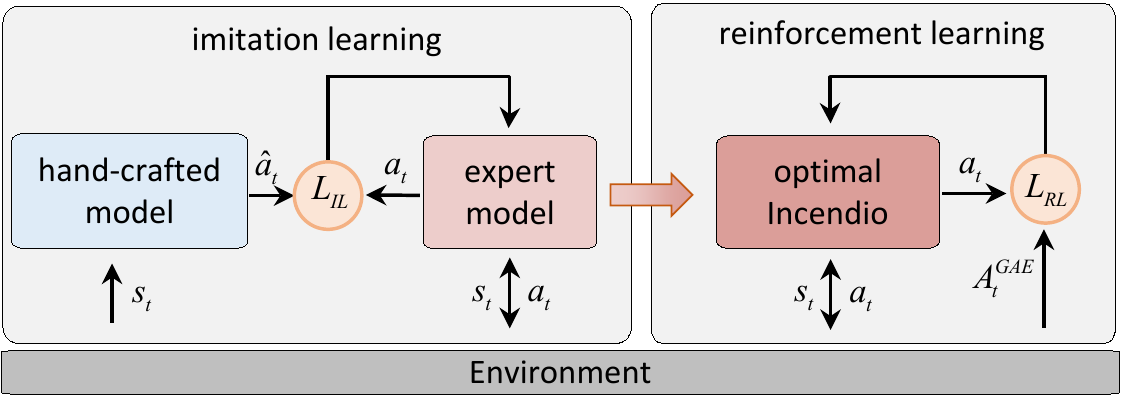}
    \caption{{\it Two-stage training} of Incendio: it is first initiated by imitating the hand-crafted model and then fine-tuned with reinforcement learning.}
    \label{fig:train}
\end{figure}

\section{Incendio Design}
As depicted in Figure~\ref{fig:train}, Incendio's training process comprises of two stages. In the first stage, we pre-train the Incendio's two agents individually by imitating a hand-crafted expert policy, which prevents them from massive inefficient explorations in the early training. Subsequently, Incendio's policy is further fine-tuned using multi-agent reinforcement learning (MARL) until converging to a global optimum. Notably, Incendio utilizes a centralized training and decentralized execution (CTDE) approach to train Incendio's two agents, in which they collaborate to attain a shared reward objective. This training strategy not only facilitates the efficient policy update for Incendio but also avoids it from converging to sub-optimal policies. 

This section first introduces the input states, actions, and neural networks of Incendio's agents, which remain consistent across the pre-train and fine-tune training stages. Then the training algorithms for pre-training and fine-tuning are elaborated. 

\vspace{-5pt}
\subsection{Multi-agent Design}
\label{ssec:agents design}
Incendio's multi-agent takes an action ${{a}_{t}}$ based on the observations collected by clients as input state ${{s}_{t}}$, according to its policy ${{\pi }_{\theta }}({{s}_{t}},{{a}_{t}})$ which is represented by neural networks. This subsection expounds on the specification of Incendio's state, action, and neural network design.


{\bf State.} The state gathered by Incendio at step $t$ is defined as $s_t=(\vec{{{b}_{t}}},\vec{{{l}_{j}}},\vec{{g}_{j}},\vec{{{u}_{j}}},\vec{{{h}_{j}}},\vec{{q}_{j}},\vec{{f}_{j}})$. The first component is a vector of throughput measurements observed in past $K$ chunks (e.g., $\vec{{{b}_{t}}}=\{{{b}_{t-K+1}},\ldots ,{{b}_{t}}\}$), each of which can be calculated by dividing the chunk size by the download duration. The remaining state components represent different types of factors for each video $j\in [1,5]$ in the queue. Specifically, ${{{l}_{j}}}$ is the conditional probability that a user will continue to watch the video $j$ from the current playing chunk $m$ (for recommended video, $m=1$) until the chunk $n$, which can be calculated as follows:
\begin{equation}
\label{eq:cond_prob}
    \begin{aligned}
        l_{j}=\frac{p_j^n}{p_j^m},
    \end{aligned}
\end{equation}
where ${p_j^m}$ denotes the user retention rate of video $j$ in chunk $m$ by statistical averaging; ${{g}_{j}}$ represents the current buffer size for video $j$; ${{{u}_{j}}}$ denotes the mean size of next chunks at different bitrates for video $j$; ${{{h}_{j}}}$ is the rebuffering time caused by downloading the last chunk for video $j$ and is equal to $0$ if no rebuffering occurred; and ${{q}_{j}}$ and ${{f}_{j}}$ respectively denote the bitrate and its fluctuation at which the last chunk was downloaded for video $j$. The bitrate fluctuation can be obtained by
\begin{equation}
\label{eq:br_smooth}
    \begin{aligned}
        f_{j}= \left|{q_j} - {q_{j - 1}}\right|.
    \end{aligned}
\end{equation}

The state for BM-agent is defined as $s_t^{bm}=(\vec{{{b}_{t}}},\vec{{{l}_{j}}},\vec{{g}_{j}},\vec{{{u}_{j}}})$. And the state for BA-agent to make the decision on video $j$ is defined as $s_t^{ba}=(\vec{{{b}_{t}}},{{{l}_{j}}},{{g}_{j}},{{{u}_{j}}},{{{h}_{j}}},{{q}_{j}},{{f}_{j}})$. In this work, we set $K = 5$ empirically to capture the temporal features from past observations.

{\bf Action.} The output action $a_t^{bm}$ of BM-agent is a 0-1 vector with a length of $6$. The first five values in this vector represent the corresponding video ID, while the last value signifies sleep for a fixed duration of $\tau=200ms$. This setting is motivated by the need to balance the trade-off between utilizing computing resources optimally and not missing the ideal decision-making time. Similarly, the BA-agent takes an action $a_t^{ba}$ from the bitrate ladders of $\{750,1200,1850\}$ kbps (same as MMGC2022), which correspond to different video qualities.

\begin{figure}[t]
    \setlength{\abovecaptionskip}{0.1cm}
    \centering
    \setlength{\belowcaptionskip}{-12pt}
    \includegraphics[width=\linewidth]{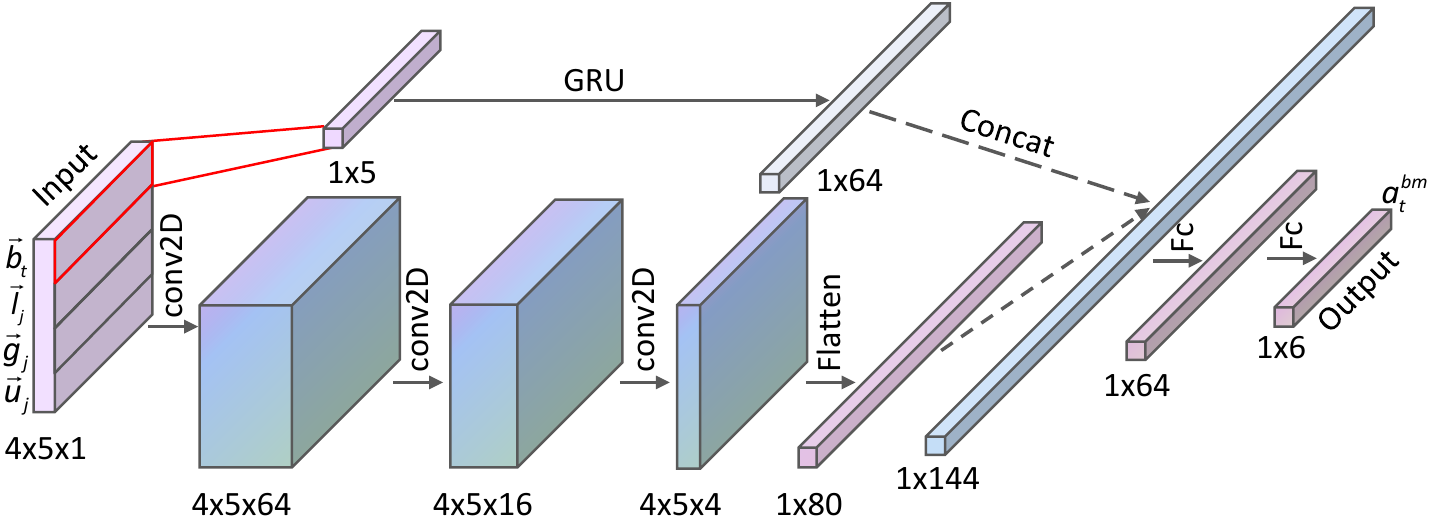}
    \caption{The NN architecture of Incendio's BM-agent.}
    \label{fig:BMagent}
\end{figure}

\textbf{Neural networks.} 
Incendio employs the actor-critic framework to represent its control policy. The actor network structure for the BM-agent is depicted in Figure~\ref{fig:BMagent}. To capture temporal features from the past bandwidth, we adopt a gated recurrent unit (GRU) layer with 64 units. Moreover, to extract spatial features between input vectors, we use three 2-D convolutional network layers with kernel sizes of 5x5 and output channels of 64, 16, and 4, respectively. The output is then concatenated and passed through a fully connected layer with 64 units. Each layer of the network uses leaky-ReLU as the activation function. At the output layer, we use the SoftMax activation function in a fully connected network to obtain a 6-dimension vector that represents the probabilities of choosing each action. For BA-agent’s actor network, we substitute the convolution layers with a fully connected layer, while keeping the remaining parts identical to the BM-agent. BM-agent and BA-agent share one critic network, which concatenates the last hidden layer of BM-agent's actor network and that of BA-agent's to output a tensor as \textit{value} without activation function. This well-designed neural network enables Incendio to efficiently extract temporal and spatial features from observations and eventually achieve excellent performances.

\vspace{-5pt}
\subsection{Pre-train with IL}
\label{ssec:IL Training}
Imitation learning is a type of machine learning where an agent learns to perform a task by observing trajectories produced by an expert and has proved to be effective in various fields, e.g., robotic, autonomous vehicles, and network streaming~\cite{10.1145/3054912, zhang2016query,comyco}. The reason we use imitation learning to pre-train Incendio is that it can reduce the amount of trial and error needed for an agent to learn a task, leading to faster convergence and more efficient learning. 

{\bf Expert policy.} The quality of expert policy directly determines the height that an agent can achieve via imitation learning. Inspired by a recent research on combining the learning-based method with a rules-based method to further improve the performance~\cite{loki}, we use the state-of-the-art hand-crafted method PDAS~\cite{PDAS} to guide Incendio. As detailed in §\ref{sec:backgrounds}, PDAS integrates user retention rate to enhance the accuracy of QoE estimation, and employs MPC techniques to enable an optimal decision-making process by exhaustively analyzing all potential future decision combinations, thus attaining good performance. 
In this work, we estimate the QoE of each possible action combination using the model designed by PDAS based on the real throughput over a horizon of future chunks and pick the largest one as the expert policy. Then we separate this expert policy into two subsets: the video ID trajectory and the bitrate trajectory, which are individually used to train Incendio's two agents respectively. Please refer to PDAS for more details. {Note that PDAS is mainly used to improve the efficiency of the exploration at the early stage of the training, and can be replaced by any other outstanding rule-based method.}

{\bf Loss function.} Similar to traditional supervised learning where samples consist of feature-label pairs, imitation learning is characterized by the demonstration of state-action pairs. Therefore, the cross-entropy function, which is widely used in classification problems, also applies here. The loss function of imitation learning for Incendio is described as follows:
\begin{equation}
\setlength{\belowdisplayskip}{-1pt}
\label{eq:il_loss}
    L_{IL} =-\sum_{t} \hat{A}_{t} \log\pi_{\theta} (s_t,a_t),
\end{equation}
where $\pi_{\theta} (s,a)$ is the policy of the agent with parameter $\theta$. $\hat{A}$ is the action probability list generated by expert policy, where the value of expert action $\hat{a}$ is equal to 1 and the others are 0. We use the policy gradient method~\cite{NIPS1999_464d828b} as the training strategy for both IL and RL. Its main idea is to estimate the gradients of the expected total reward with respect to the policy parameter $\theta$, and update the network parameters according to the gradients, which can be written as follows:
\begin{equation}
\label{eq:grad_update}
    d \theta \leftarrow d \theta-\alpha \sum_{t} \nabla_{\theta} \hat{A}_{t} \log \pi_{\theta}(s_t,a_t).
\end{equation}
We set the learning rate $\alpha=0.0001$. The BM-agent and BA-agent of Incendio are trained with the same loss function. We also introduce a novel experience replay mechanism to improve the sample utilization efficiency.

\vspace{-5pt}
\subsection{Fine-tune with RL}
\label{ssec:RL Training}
Incendio fine-tunes its SABR policy using a state-of-the-art MARL framework, i.e., multi-agent proximal policy optimization (MAPPO~\cite{mappo}) algorithm which is an effective improved version of PPO~\cite{ppo} designed for multi-action tasks. At each interaction step, Incendio takes an action ${{a}_{t}}$ including ${{a}_{t}^{bm}}$ and ${{a}_{t}^{ba}}$ according to its policy ${{\pi }_{\theta }}$ and the observations ${s}_{t}$. Then the environment transits to a new state $s_{t+1}$ and returns a reward $r_t$ which will be used to update the NN parameters of both agents for Incendio. From the perspective of one agent, changes brought about by another agent will be regarded as environmental changes.

{\bf Reward.} In a typical RL task, the agent learns the optimal policy by maximizing the expected cumulative (discounted) reward that it receives from the environment. Thus, we set the reward to reflect the MMGC2022 utility score that is defined in Equation~\eqref{eq:score}. Aiming to maximize the expected cumulative discounted reward, Incendio's RL agents learn to get higher utility scores. To motivate Incendio to learn a more bandwidth-efficient policy, we add a coefficient $w_n$ to the first term of the reward function which could be formulated as follows:
\begin{equation}
\label{eq:reward}
    r_{t} = w_n \cdot l_n \cdot \left(R_{n}-S_{n}\right)- \mu \cdot T_{n}- \nu \cdot bw_{n},
\end{equation}
where $l_n$ is the conditional probability as defined in Equation~\eqref{eq:cond_prob}. The idea behind this is that if the user retention rate of the next to-be-downloaded chunk is much lower than that of the current playing one, which means there is a high probability that the user will swipe away before playing the next chunk, and the agent will receive a lower reward. To get a higher reward, the agent will try to pause the download until the conditional probability increase, leading to less bandwidth wastage. $w_n$ indicates whether the last downloaded chunk $n$ will be watched by the user and is defined as follows:
\begin{equation}
\label{eq:w_n}
    w_{n} = \begin{cases}
  0,& \text{ if user will scroll to next video before chunk } n \\
  1,& \text{ if user will watch chunk } n.
\end{cases} 
\end{equation}
In the training, $w_n$ is sampled randomly in the user retention rate trace whenever a video is added to the video queue. During the evaluation, we still use Equation~\eqref{eq:score} to evaluate the performance.

\begin{figure*}[t]
    \centering
    \subfigcapskip = -8pt
    \abovecaptionskip = -1pt
    \setlength{\belowcaptionskip}{-10pt}
    \subfigure[Average results]{
    \begin{minipage}[t]{.63\linewidth}
        \centering
        \includegraphics[width=\linewidth]{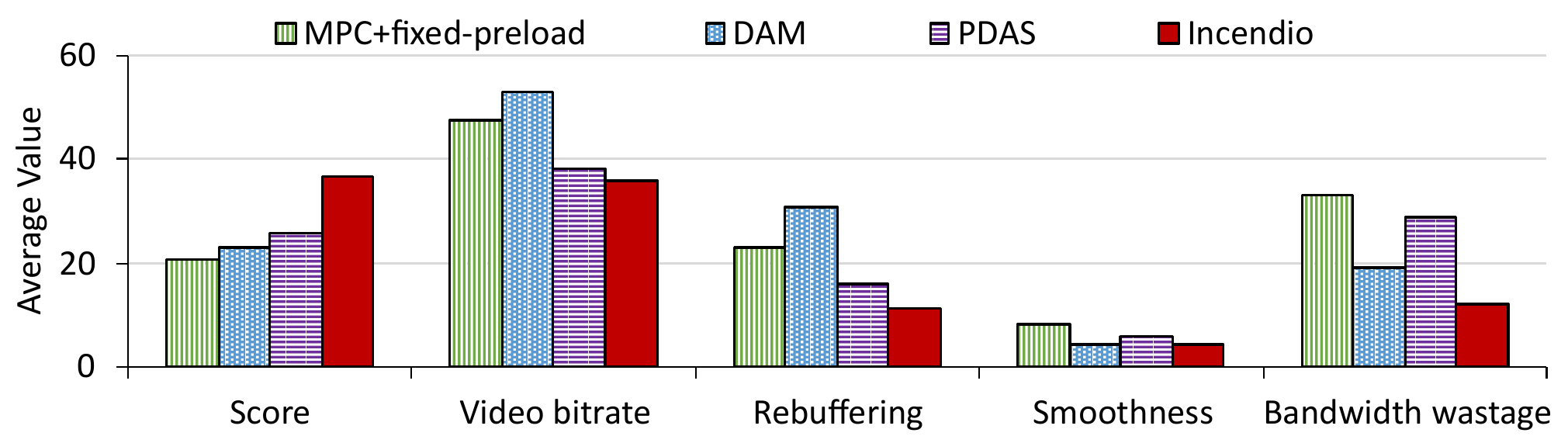}
        \label{sfig:results_metrics}
    \end{minipage}
    }
    \subfigure[CDF results]{
    \begin{minipage}[t]{.3\linewidth}
        \centering
        \includegraphics[width=\linewidth]
        {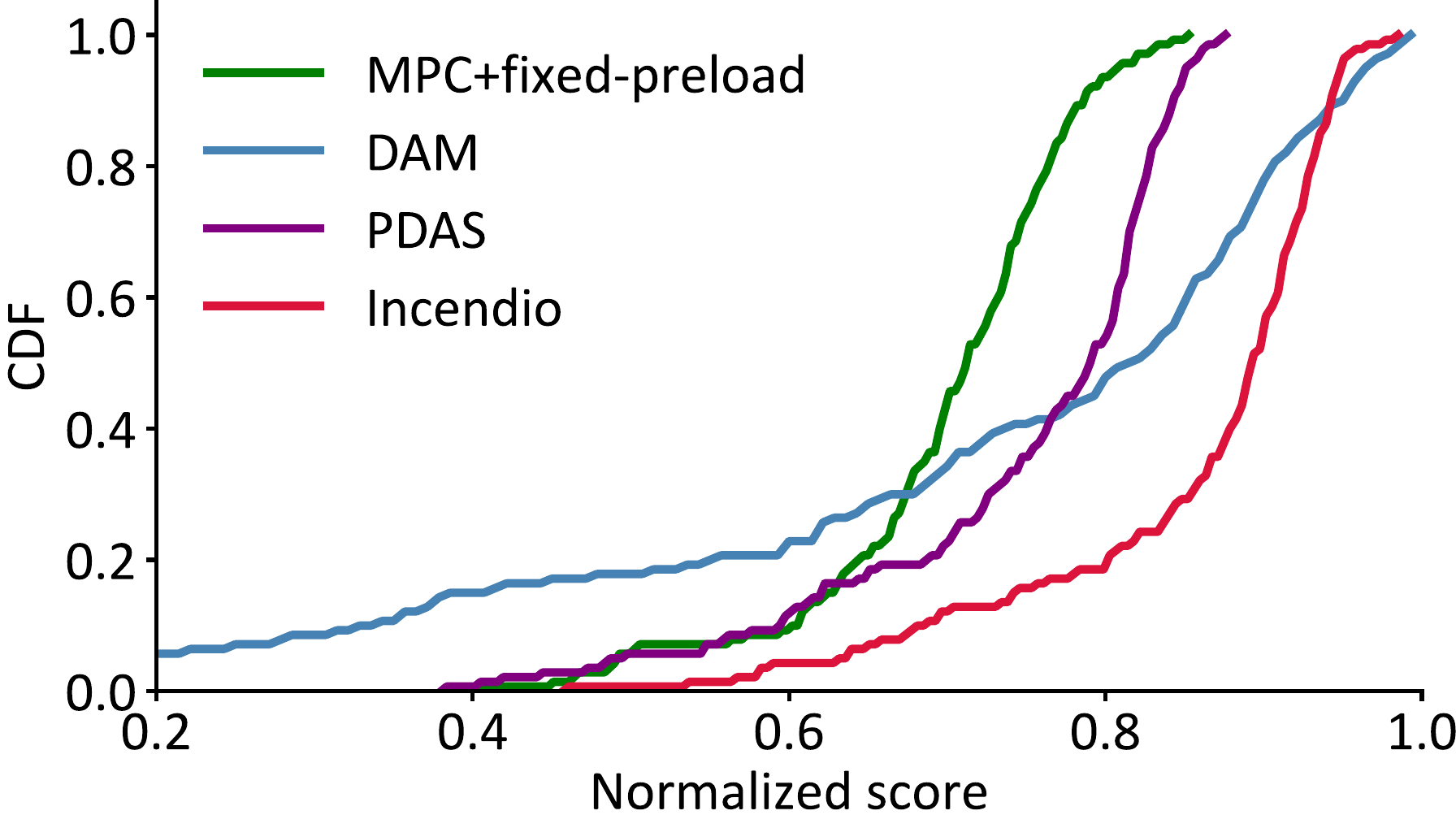}
        \label{sfig:result_cul}
    \end{minipage}
    }
    \caption{Comparing Incendio with the other schemes in terms of the average performance and full CDF performance under MMGC2022 video dataset and bandwidth trace dataset. The scores are normalized for the CDF results.}
    \label{fig:results}
\end{figure*}

\begin{figure*}[t]
    \centering
    \subfigcapskip = -8pt
    \abovecaptionskip = -1pt
    \setlength{\belowcaptionskip}{-10pt}
    \subfigure[Average results]{
    \begin{minipage}[t]{.63\linewidth}
        \centering
        \includegraphics[width=\linewidth]{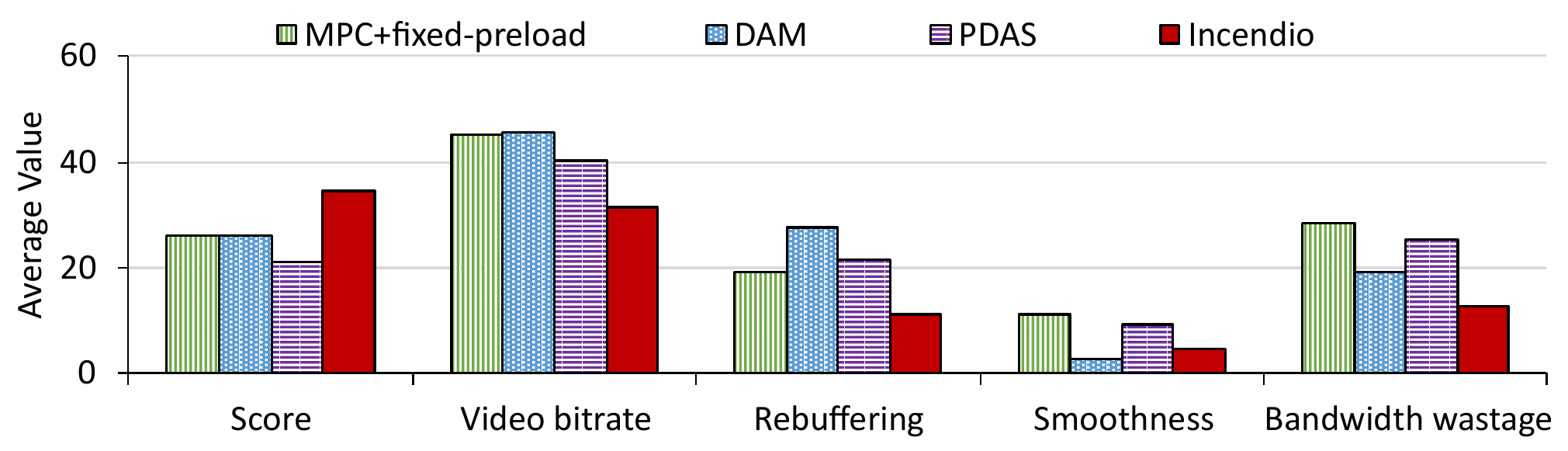}
        \label{sfig:results_metrics_oboe}
    \end{minipage}
    }
    \subfigure[CDF results]{
    \begin{minipage}[t]{.3\linewidth}
        \centering
        \includegraphics[width=\linewidth]{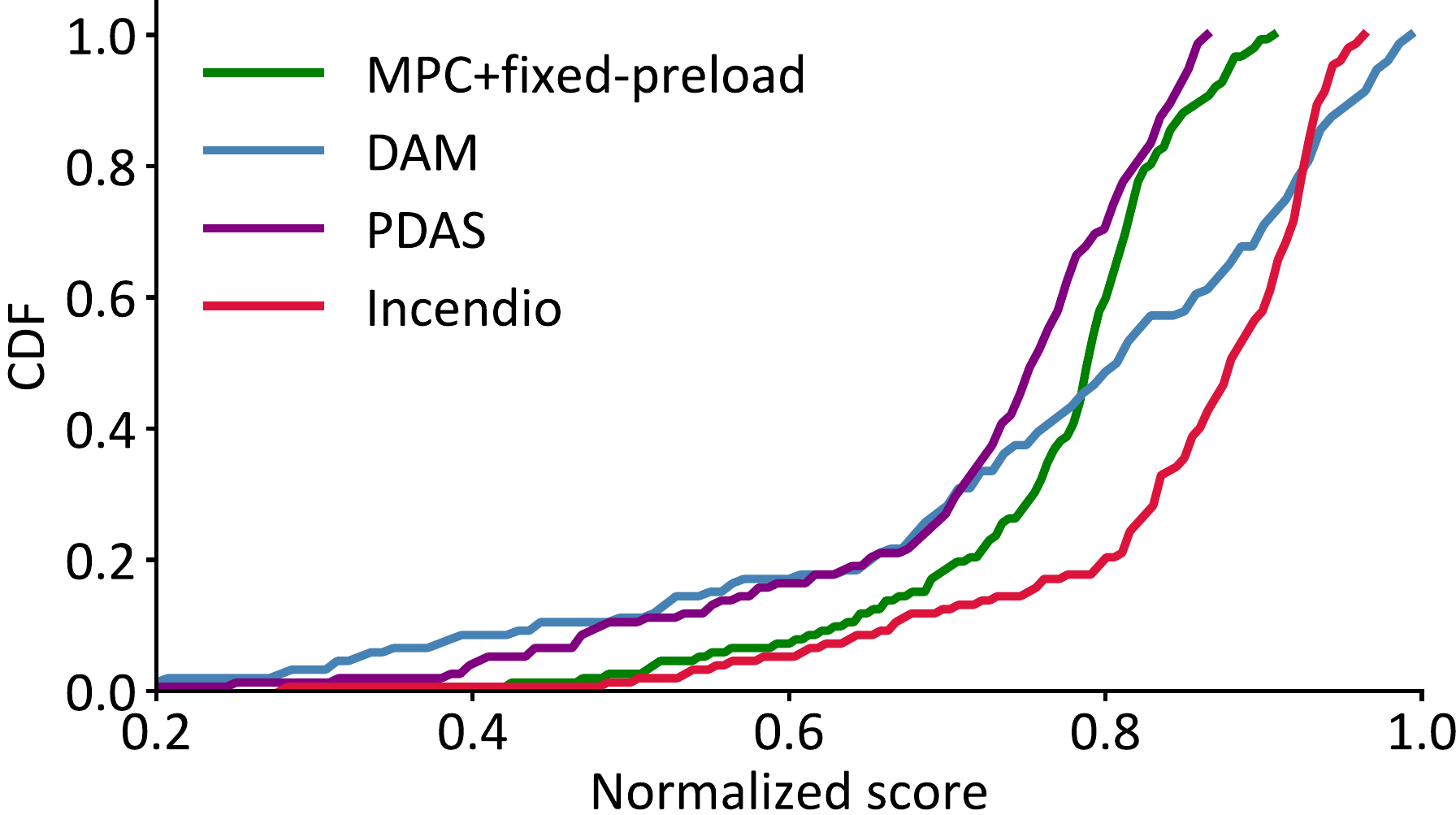}
        \label{sfig:result_cul_oboe}
    \end{minipage}
    }
    \caption{Comparing Incendio with the other schemes in terms of the average performance and full CDF performance under DUASVS video data set and Oboe/FCC bandwidth trace datasets. The utility scores are normalized for the CDF results.}
    \label{fig:results_oboe}
\end{figure*}

\textbf{Training methodology.} We use the clipped surrogate loss function to train Incendio's RL agents and additionally introduce the entropy of policy to avoid converging to sub-optimal policies at the early stage of training. The loss function is formulated as follows:
\begin{equation}
    \resizebox{0.92\hsize}{!}{$\begin{aligned}
    L_{RL}= & -\sum_{t} \sum_{k} \min \left(ratio_{\theta ,t}^{k}, \operatorname{clip}\left(ratio_{\theta, t}^{k}, 1-\epsilon, 1+\epsilon\right)\right)A_{t}^{GAE} \\ & +\beta\sum_{t} \sum_{k} H\left(\pi_{\theta}\left(s_{t}^{(k)}\right)\right).
    \end{aligned}$}
\end{equation}
Here $A_t^{GAE}$ is the {\it advantage function} computed using the GAE~\cite{gae} method, which represents the difference in the expected reward when the agent deterministically picks action $a_t$ in state $s_t$, compared with the expected reward for actions following the policy ${{\pi }_{\theta }}$ with the policy parameters $\theta$. $ratio_{\theta, t}^{k}$ and $H(\cdot)$ represent the surrogate objective and policy entropy respectively. k indexes the agents where $k\in\left \{ bm,ba \right \}$. $\beta$ is the weight of the entropy term and we decay it when the reward does not increase for 100 epochs. The gradient update formula is similar to Equation~\eqref{eq:grad_update} and more technical details with respect to the training algorithm can be found in~\cite{mappo}.



\section{Evaluation}
\subsection{Methodology}
\label{ssec:Methodology}
To evaluate the performance of Incendio, we utilize the multi-video simulator provided in MMGC2022 to simulate various short video streaming sessions by randomly combining different video traces and network traces.  As for video traces, each chunk size at different bitrates is recorded in a video size trace track, and corresponding user retention rates per chunk are contained in a user retention rate trace track. 

To train Incendio, we create a corpus of network traces by combining some public datasets including Oboe~\cite{Oboe} and FCC~\cite{fcc2016}. As for video traces, we use the DUASVS~\cite{DUASVS} which contains millions of records including the video chunk statistics and users' retention rates. Unless otherwise noted, we used a random sample of 80\% of our corpus as a training set for Incendio and the remaining 20\% as a testing set for all SABR algorithms. The same training set is also used to train DAM for a fair comparison. Since both network and video traces in MMGC2022 are relatively small, we only use this MMGC2022 dataset for evaluation. Note that the network traces provided by MMGC2022 record bandwidth samples over time under high, medium, and low network conditions, respectively.

We compare Incendio to the following schemes, which collectively represent the state-of-the-art methods:
\begin{itemize}[leftmargin=*]

\item {\it MPC+fix-preload}, combines a prevalent model-based ABR approach (i.e., RobustMPC~\cite{MPC}), with the sleeping mechanism used in {\it fix-preload}~\cite{mmgc2022}, which is the baseline provided by MMGC2022. RobustMPC maximizes the accumulated utility score defined by Equation~\eqref{eq:score} over a horizon of 5 future chunks based on the buffer occupancy observations and throughput predictions. 
\item {\it DAM}~\cite{DAM}, a deep reinforcement learning-based approach, which trains a control policy to make the decision of video ID, bitrate level, and pause/sleep time jointly using an action masking mechanism. It ranks first among all learning-based techniques in MMGC2022. We faithfully implement DAM following its paper.
\item {\it PDAS}~\cite{PDAS}, the state-of-the-art approach in MMGC2022 which optimizes the utility score by jointly utilizing user retention rate and a handcrafted buffer management model. PDAS employs RobustMPC to enable optimal decision-making. We faithfully implement PDAS following its paper.
\end{itemize}

\vspace{-8pt}
\subsection{Overall Results}
\label{ssec:Results}

Figure~\ref{fig:results} and Figure~\ref{fig:results_oboe} demonstrate the results of performance comparison between Incendio and other schemes in terms of average utility score metrics and related full cumulative distribution function (CDF) under different short video datasets (MMGC2022 and DUASVS) and bandwidth datasets (MMGC2022 and Oboe/FCC). As clearly reported in Figure~\ref{sfig:results_metrics} and Figure~\ref{sfig:results_metrics_oboe}, Incendio gains a clear leading position on the metric of both average score and some individual components (i.e., rebuffering, smoothness, and bandwidth wastage). Specifically, Incendio outperforms the state-of-the-art PDAS by 53.2\% on the average utility score. Furthermore, Incendio achieves reduction of 39.1\% - 61.8\% on rebuffering and 34.9\% - 59.2\% on bandwidth wastage, which are remarkable improvements compared to other schemes. The results in the form of full CDF, shown in Figure~\ref{sfig:result_cul} and Figure~\ref{sfig:result_cul_oboe}, further demonstrate the consistent performance of Incendio.

As illustrated in Figure~\ref{sfig:results_metrics}, PDAS outperforms MPC+fixed-preload for nearly all the metrics, indicating  that it is a more advanced variant of MPC. However, PDAS shows inferior performance to DAM in bandwidth saving and managing bitrate fluctuation, suggesting that the max buffer model of PDAS is not well-designed. On the other hand, PDAS reports the worst performance using the DUASVS video data set under the Oboe/FCC network conditions, which reveals the poor generalization of PDAS. We believe that this is because the hyperparameters in PDAS fine-tuned using the MMGC2022 dataset are not able to characterize the network dynamics of Oboe/FCC traces and the variation of user preference in the DUASVS video set.


Interestingly, we find that DAM demonstrates unstable performance at different QoE ranges as visualized in Figure~\ref{sfig:result_cul} and Figure~\ref{sfig:result_cul_oboe}, reporting relatively higher performance within the high QoE range ($>$ 0.8 approximately) but the noticeable lower performance in the low and intermediate QoE ranges ($<$ 0.8) compared to other competitors. The reason is that DAM tends to discard certain actions to facilitate the exploration process due to the vast multidimensional exploratory space. Furthermore, we find evidence of the above analysis from the training log of DAM: it always chooses the middle-level bitrate for recommended video and never selects the lowest bitrate for the current playing video. The metric performance of DAM is also consistent with our findings, whereby being partially owed to choosing high-quality (high bitrate) chunks leads to higher probabilities of rebuffering. 

\begin{figure}[t]
    \setlength{\abovecaptionskip}{0.1cm}
    \centering
    \setlength{\belowcaptionskip}{-12pt}
    \includegraphics[width=0.9\linewidth]{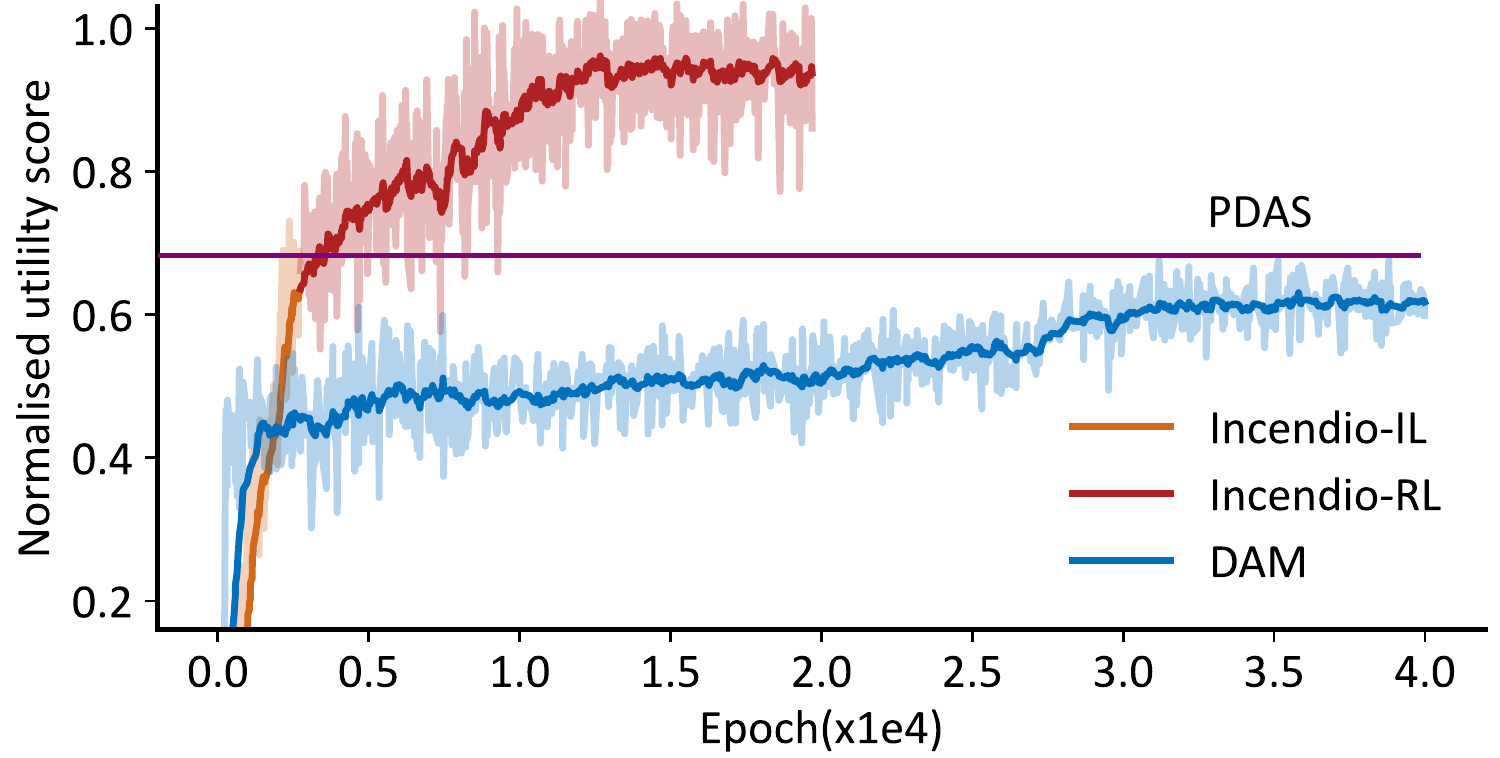}
    \caption{The training log of Incendio and DAM.}
    \label{fig:log}
\end{figure}

\subsection{Training Efficiency}
\label{ssec:convergence}
We plot the training log of Incendio and DAM in Figure~\ref{fig:log} to compare the training efficiency of our two-stage training approach and the centralized RL method of DAM. As shown, DAM falls into a sub-optimal policy at the early stage of the training and takes a long convergence time duration to the final policy. That's because DAM encounters a significant challenge due to the substantial exploration space generated by all possible combinations of atomic actions. In contrast, Incendio demonstrates the ability to learn an expert-level policy within a significantly shorter duration of 4k epochs via imitation learning. Furthermore, our proposed two-stage training framework enables Incendio to make substantial progress during the reinforcement learning phase. These results demonstrate the efficiency and significance of our training approach.

\subsection{Inference Time}
\label{ssec:running_time}
To evaluate the complexity of Incendio, we record its runtime in the inference phase and compare it with that using MPC+fix-preload, DAM, and PDAS. We set two different environments for evaluation. In the environment $E1$, we keep the same setting as in the MMGC2022 competition, while the numbers of videos and bitrate levels are slightly increased to 7 and 6 respectively in the $E2$. The experiments are conducted on a desktop equipped with an Intel(R) Core(TM) i5-12500@3.00GHz CPU and repeated thousands of times. Table~\ref{tab:inference_time} lists the average inference times for MPC+fix-preload, DAM, PDAS, and Incendio. As shown, the computation complexity of MPC+fix-preload and PDAS grows exponentially as the numbers of videos in the queue and bitrate levels increase, which is unacceptable for mobile devices. This is because these two MPC-based SABR algorithms make decisions by comparing the expected rewards of all possible combinations of future actions, which limits its application in scenarios with large action space. On the contrary, Incendio maintains a lightweight computational complexity in these environments, which rivals DAM at the same order of magnitude but provides superior SABR performance (§\ref{ssec:Results}) and higher training efficiency (§\ref{ssec:convergence}). 


\begin{table}[t]
\setlength{\abovecaptionskip}{0.1cm}
\caption{The inference time for Incendio and existing SABR algorithms in two different environments. Time is measured by milliseconds.}
\label{tab:inference_time}
\begin{tabular}{c|c|c|c|c}
\hline
       & MPC+fixed-preload & DAM  & PDAS   & {\bf Incendio}   \\ \hline
E1     &    0.6           & 0.4  &  7.1 & {\bf 1.0}   \\
E2     &   11.8             & 0.4  & 318.1    & {\bf 1.1}       \\ \hline
\end{tabular}
\vspace{-5pt}
\end{table}

\vspace{-5pt}
\section{Conclusion}
We proposed and evaluated Incendio, a novel ABR framework for short video streaming using multi-agent reinforcement learning with expert guidance, with which we separate the decision of buffer management (e.g., video prefetching) and bitrate adaptation in a sequential manner to optimize the system-level utility score. As a result,  the training of Incendio is initiated with human expert policy using Imitation learning and then uses a much smaller action space for policy fine-tuning, providing a much faster convergence rate and exhibiting exceptional generalization abilities. Over a broad set of testing conditions, we find the proposed Incendio provides more than 2x performance improvement to the state-of-the-art scheme when using the utility score measurement.


\vspace{-5pt}
\begin{acks}
This work is partially supported by National Natural Science Foundation of China (62101241), Jiangsu Provincial Double-Innovation Doctor Program (JSSCBS20210001).
\end{acks}

\bibliographystyle{ACM-Reference-Format}
\bibliography{sample-base}

\end{sloppypar}
\end{document}